\documentclass[%
 reprint,
 amsmath,amssymb,
 aps,
floatfix,
]{revtex4-2}

\usepackage{graphicx}
\usepackage{dcolumn}
\usepackage{bm}

\usepackage{xcolor}
\usepackage{amsmath}
\usepackage{dsfont}
\newcommand{\beq}{\begin{equation}}
\newcommand{\eeq}{\end{equation}}
\newcommand{\ud}{\mathrm{d}}
\newcommand{\kk}{\mathbf{k}}

\begin{document}

\preprint{APS/123-QED}

\title{High-Field Superconducting Halo in UTe$_2$}

\author{Sylvia K. Lewin$^{1,2}$}
\thanks{These authors contributed equally: Sylvia K. Lewin and Peter Czajka}

\author{Peter Czajka$^{1,2}$}%
\thanks{These authors contributed equally: Sylvia K. Lewin and Peter Czajka}

\author{Corey E. Frank$^{1,2}$}

\author{Gicela Saucedo Salas$^2$}

\author{Hyeok Yoon$^2$}

\author{Yun Suk Eo$^2$}

\author{Johnpierre Paglione$^{2,3}$}

\author{Andriy H. Nevidomskyy$^4$}

\author{John Singleton$^5$}

\author{Nicholas P. Butch$^{1,2}$}
 
\affiliation{$^1$NIST Center for Neutron Research, National Institute of Standards and Technology, 
Gaithersburg, MD, USA}

\affiliation{$^2$Maryland Quantum Materials Center, Department of Physics, University of Maryland, College Park, MD, USA}

\affiliation{$^3$Canadian Institute for Advanced Research, Toronto, Ontario M5G 1Z8, Canada}

\affiliation{$^4$Department of Physics and Astronomy, Rice University, Houston, TX, USA}

\affiliation{$^5$National High Magnetic Field Laboratory, Los Alamos National Laboratory, Los Alamos, NM, USA}

\date{\today}

\begin{abstract}
Heavy fermion UTe$_2$ is a promising candidate for topological superconductivity that also exhibits multiple high-field superconducting phases. The SC$_{\rm{FP}}$ phase has only been observed in off-axis magnetic fields in the $bc$ plane at fields greater than 40 teslas, a striking scale given its critical temperature of only 2 kelvins. Here, we extend measurements of this unique superconducting state outside of the $bc$ plane and reveal its core structure. The SC$_{\rm{FP}}$ phase is not confined to fields in the $bc$ plane and in fact wraps around the $b$ axis in a halo-like fashion. In other words, this superconducting state, which exists in fields above 73 teslas, is stabilized by a field component perpendicular to the magnetic easy axis. 
These remarkable field scales further underscore UTe$_2$'s unique magnetophilic superconducting tendencies and suggest an underlying pairing mechanism that is qualitatively distinct from known theories for field-enhanced superconductivity. Phenomenological modeling points to a two-component, non-unitary spin triplet order parameter with finite orbital momentum of the Cooper pairs  as a natural explanation for the field-angle dependence of the upper critical field of the SC$_{\rm{FP}}$ phase.
\end{abstract}

\maketitle


Magnetic fields typically act to destroy superconductivity, a fact that places strict physical limitations on superconducting technologies. Heavy fermion material UTe$_2$ represents a striking exception to this conventional phenomenology \cite{ran2019nearly,ran2019extreme,lewin2023review,aoki2022unconventional}. In addition to harboring a low-field superconducting state that is of interest for topological quantum computing applications \cite{jiao2020chiral,hayes2021multicomponent}, the material also exhibits additional superconducting phases that only exist at extraordinarily high magnetic fields, as illustrated in the full phase diagram in Fig. 1a. These include a field-reinforced state that exists up to 34 T for \textbf{H} aligned close to the UTe$_2$ $b$ axis (SC2)\cite{ran2019extreme,knebel2019field} as well as an even more exotic phase that has only been observed by tilting \textbf{H} approximately 20-40$^o$ from $b$ towards $c$ (SC$_{\rm{FP}}$) \cite{ran2019extreme}. The latter, which is the focus of the present study, only exists within the field-polarized (FP) state that is separated from the low-field regime by a metamagnetic transition (pink sheet in Fig. 1a). Remarkably, the SC$_{\rm{FP}}$ phase only onsets above 40 T and persists up to at least 73 T. These field scales are particularly impressive given UTe$_2$'s relatively low $T_{\rm{c}}$ (1.6-2.1 K depending on sample) \cite{ran2019nearly, sakai2022single}. We also emphasize that the material is fairly three-dimensional \cite{YunSuk,Thebault}, which requires the superconductivity to be robust against orbital as well as spin-based depairing mechanisms, unlike the numerous Pauli limit-breaking 2D systems for which high critical fields are only seen when \textbf{H} is directed within the conducting plane \cite{lu2015evidence,singleton2002quasi,Balicas}. It also differs from other uranium-based superconductors for which field-enhanced pairing is more straightforwardly associated with specific phase boundaries and high-symmetry directions \cite{aoki2019review,levy2005magnetic}.

\begin{figure*}
    \centering
    \includegraphics[width=\textwidth]{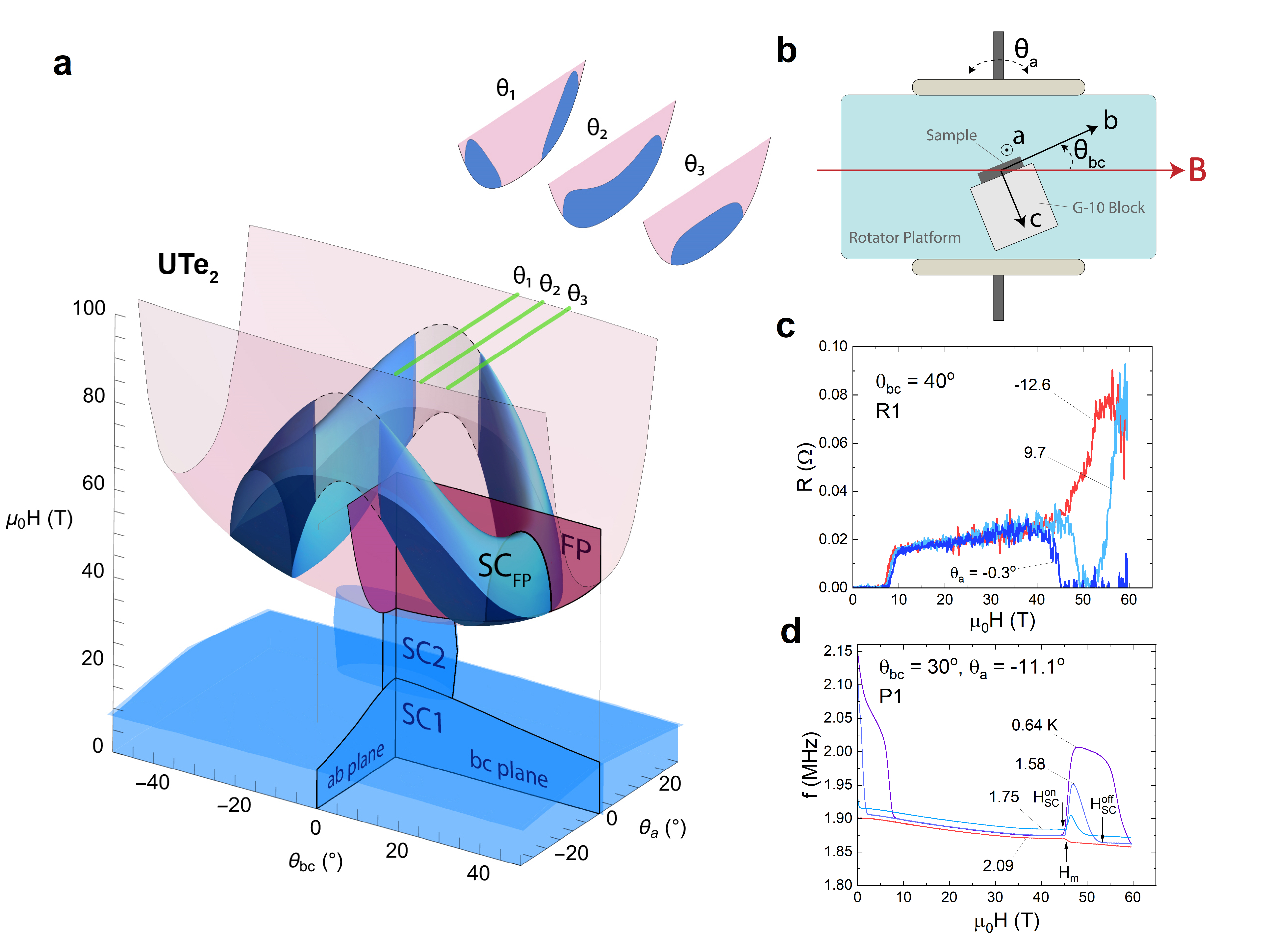}
    \caption{\textbf{Field-angle phase diagram of UTe$_2$ and experimental details.} (a) Illustration of UTe$_2$'s field-dependent phase diagram, with the $ab$ and $bc$ planes highlighted. The three superconducting regions are shown in blue. The boundaries of the low-field SC1 phase vary smoothly with the field direction, while the field-stabilized SC2 phase exists only for fields near the $b$ axis. The SC$_{\rm{FP}}$ state appears inside the field-polarized state (pink) only for specific off-axis fields that yield the saddle-like shape in the drawing. (b) Depiction of measurement setup that enables effective dual-axis rotation. The crystal (gray) is placed on a block (white) on the rotator platform at a fixed  angle within the $bc$ plane ($\theta_{bc}$); the angle off the $bc$ plane ($\theta_{a}$) is set to different values during the experiment. (c) Electrical resistance $R$ vs. field at various $\theta_a$ for sample R1, which reveal the high-field zero-resistance SC$_{\rm{FP}}$ state. (d) PDO data obtained at various temperatures for sample P1. An abrupt jump in $f$ (indicating a transition to a low resistance state) is no longer observed at 2.09 K, consistent with a loss of superconductivity at that temperature.}
    \label{fig:1}
\end{figure*}

Prior experiments on the SC$_{\rm{FP}}$ phase have been limited to crystallographic planes. Superconductivity was found with magnetic fields applied in the $bc$ plane \cite{ran2019extreme,knafo2021comparison}, but not the $ab$ plane \cite{ran2019extreme}, which has fueled speculation that there may be some special angle within the $bc$ plane that permits this unique high-field phase \cite{knafo2021comparison}. In this study, we map the SC$_{\rm{FP}}$ phase boundaries outside the $bc$ plane and find that SC$_{\rm{FP}}$ actually wraps around $b$ in a halo-like fashion (high $H$ blue region in Fig. \ref{fig:1}a). We emphasize that Fig. \ref{fig:1}a is an illustration intended to capture broad features in an extended data set.

All measurements were performed at the National High Magnetic Field Laboratory's Pulsed Field Facility at Los Alamos National Laboratory. Data were primarily taken in a 65 T magnet system (typical field pulses were 60 T) as well as a 75 T duplex magnet (typical field pulses were 73 T). Lacking a dual axis rotator suitable for pulsed fields, we utilized a single axis rotator with the crystal manually placed on the sample platform such that \textbf{H} is applied at some angle $\theta_{bc}$ within the bc plane (see Fig. \ref{fig:1}b). The rotator is then used to apply field pulses at various tilt angles $\theta_a$ towards the UTe$_2$ $a$ axis. An example of this for $\theta_{bc}$ = 40$^o$ is shown in Fig. \ref{fig:1}c. A high-field zero resistance state is obtained when \textbf{H} is applied within the $bc$ plane ($\theta_a$ $\approx$ 0$^o$). As \textbf{H} is tilted towards $a$, the $H$ window in which superconductivity is observed shrinks until it is lost entirely around $\theta_a =$ 12.6$^o$. For $\theta_a$ $\geq$ 12.6$^o$ the resistance instead shows a sharp increase, which is a signature of the metamagnetic transition \cite{knafo2019magnetic}.

For some samples we utilized a contactless conductivity technique: Proximity Detector Oscillator (PDO) measurements \cite{altarawneh2009proximity}. UTe$_2$'s low electrical resistivity and challenging surface chemistry can make traditional electrical resistivity measurements difficult, especially in pulsed field. PDO experiments circumvent these technical obstacles and are a well established technique for determining phase boundaries in UTe$_2$ and other superconductors \cite{ran2019extreme,ran2021expansion,broyles2023revealing,wu2023enhanced,sebastian2012towards,nikolo2017upper,smylie2019anisotropic}. This is illustrated in Fig. \ref{fig:1}d. At 0.64 K, dramatic shifts in PDO frequency $f$ are seen at 8 and 45 T. Higher $f$ is associated with lower resistance; the regions of elevated $f$ are superconducting. As $T$ is increased, the superconducting phase boundaries shift in the expected fashion and at 2.09 K they are no longer observed. We instead observe a kink downwards at 45 T, indicating a transition to a high-resistance FP phase. An extended discussion of PDO data analysis is provided in the Supplementary Materials.

Fig. \ref{fig:2} displays a set of field pulses and corresponding phase diagrams measured at three different $\theta_{bc}$. The core finding of this experiment is immediately apparent. As $\theta_{bc}$ decreases (the measurement plane moves closer to the \textit{ab} plane) the effect of tilting towards \textit{a} changes dramatically. For $\theta_{bc}$ = 30$^o$ we observe that increasing $\theta_a$ initially causes an enhancement of superconductivity instead of the immediate suppression seen in Fig. \ref{fig:1}c. As $\theta_a$ is increased further, the critical field peaks around 60 T and then decreases until superconductivity is lost entirely around 15.5$^o$, leaving a U-shaped region of superconductivity. The lower boundary tracks the angular evolution of $H_{\rm{m}}$ since SC$_{\rm{FP}}$ only exists within the FP phase and not in the low $H$ state.

At $\theta_{bc}$ = 23$^o$, no superconductivity is observed in the \textit{bc} plane ($\theta_a$ = 0). However, we find that tilting \textbf{H} towards \textit{a} (by 10$^o$) induces superconductivity. Superconductivity persists up to $\theta_a$ = 16$^o$, beyond which a metamagnetic transition is again observed. Similar behavior is observed at $\theta_{bc}$ = 8$^o$ (nearly in the $ab$ plane) except that the onset $\theta_a$ and $H_{\rm{SC}}^{\rm{on}}$ have been pushed even higher. This reveals a superconducting regime that only appears at 59 T and persists well above 73 T (judging by the behavior of $f$ vs. $H$ near the maximum field). It also appears that the angular window of superconductivity may narrow substantially at low $\theta_{bc}$. However, it is difficult to say with certainty since this may be related to sample heating due to P2's large size. See the Supplementary Materials for more information.

Fig. \ref{fig:3}a shows all phase diagrams measured in this experiment. This includes the PDO data presented in Fig. 2 as well as electrical resistance measurements performed on two additional samples. The electrical resistance measurements confirm the key findings from the PDO data. An initial enhancement of $H_{\rm{SC}}^{\rm{off}}$ for the SC$_{\rm{FP}}$ phase is observed at $\theta_{bc}$ = 29$^o$ and the recovery of superconductivity by $a$ axis tilting is seen at $\theta_{bc}$ = 18$^o$. Stitching phase boundaries from these various $\theta_{bc}$ measurement planes together reveals the true structure of the SC$_{\rm{FP}}$ phase (illustrated in Fig. \ref{fig:1}a). The region of superconductivity wraps around $b$ with $H_{\rm{SC}}^{\rm{on}}$ lowest within the $bc$ plane and highest as \textbf{H} approaches the $ab$ plane, reflecting the rapid increase of $H_{\rm{m}}$ with $\theta_a$. This anisotropy is responsible for the saddle-like appearance in Fig. \ref{fig:1}a.

In Fig. \ref{fig:3}b, we construct a simpler picture that captures the primary revelation of this experiment. We ignore $H$ and plot the $\theta_{bc}$ and $\theta_a$ values at which an SC$_{\rm{FP}}$ phase is observed as closed points and the values where only $H_{\rm{m}}$ is seen as open circles. Note that the data have been symmetrized to fill all four quadrants. Blue shading is added as a guide to the eye as there is some inevitable scatter in the angular phase boundaries caused by sample dependence, angle uncertainty, and differing measurement techniques. Also note that while we did not observe superconductivity within the $ab$ plane, it may just be that $H_{\rm{SC}}^{\rm{on}}$ in that plane exceeds the 73 T maximum field of the experiment. Our efforts are summarized in the Supplementary Materials. Regardless of whether the SC$_{\rm{FP}}$ truly does extend to the $ab$ plane, the curvature of the phase boundaries is unambiguous. A finite tilt of \textbf{H} off $b$ is what ultimately produces this unprecedented extreme high-field superconducting state. This contrasts with earlier investigations that pointed to a particular direction within the $bc$ plane being special. The onset angle off $b$ for superconductivity seems to be around 20$^o$ (or alternatively a perpendicular field $H_{\perp}$ of about 17 T)  with some weak dependence on direction and sample characteristics. 

Despite the apparent similarities between 
$\theta_{a}$ and $\theta_{bc}$ in their relation to superconductivity, the two directions are crystallographically  distinct. This anisotropy is apparent in the angular evolution of the FP phase boundaries \cite{ran2019extreme} and understanding of its interplay with the superconducting halo will be key for identifying the pairing mechanism and (likely triplet) order parameter. Within the $bc$ plane, $H_{\rm{m}}$ follows $H_m = H_m^b / \cos(\theta_{bc})$ where $H_m^b$ $\approx$ 34 T represents a typical value of $H_{\rm{m}}$ for \textbf{H} $\parallel$ $b$ \cite{ran2019extreme,frank2023orphan}. In other words the metamagnetic transition occurs when $H_b$ $\approx$ 34 T, regardless of $H_c$. For \textbf{H} directed outside the $bc$ plane, $H_{\rm{m}}$ instead shows a much steeper angle dependence that does not respect this field projection pattern: 
\begin{equation}
    H_m = H_m^0 + \alpha_2 \sin^2 \theta_a + \alpha_4 \sin^4 \theta_a
\end{equation}
where $H_{\rm{m}}^{\rm{0}}$ is the value of $H_{\rm{m}}$ in the $bc$ plane. It is clear that the U-shaped $H_{\rm{m}}$ envelope (which serves as a lower bound for SC$_{\rm{FP}}$) shifts to higher $H$ as $\theta_{bc}$ increases. We fit the individual $H_{\rm{m}}(\theta_a)$ curves to eq. (1) and obtain two key findings. First, the function describes the data remarkably well. Second, the fitting parameters $\alpha_2$ and $\alpha_4$ are independent of $\theta_{bc}$. The only thing that varies is the $bc$-plane onset field $H_{\rm{m}}^0$. This point is illustrated in Fig. \ref{fig:3}c where we plot $H_{\rm{m}}-H_{\rm{m}}^0$ vs. $\theta_a$ for all fixed $\theta_{bc}$ data sets and find that they all collapse on top of each other. Fitting all phase boundaries together yields values: $\alpha_2 = 94.84$ T, $\alpha_4 = 1933.62$ T. This scaling consistency indicates that tilts towards $a$ and $c$ have orthogonal effects on the FP phase, in contrast to their effects on the superconducting phase. 

Our measured bounds of the SC$_{\rm{FP}}$ state indicate that the phase must be robust against orbital depairing effects as well as spin-based ones. A likely triplet pairing state naturally accounts for the stability against paramagnetic depairing, but would not explain the orbital resilience. In 2D materials, there is no orbital motion from in-plane fields. But it is not possible to have a complete absence of induced orbital motion for all of the field directions at which SC$_{\rm{FP}}$ occurs. 

Because the SC$_{\rm{FP}}$ phase only arises within the FP state, magnetic fluctuations likely play a role in mediating the superconducting pairing.
Notably, $b$ is the likely magnetic easy axis in the FP phase \cite{miyake2019metamagnetic}. In this context, the field-angle dependence of the superconducting halo also indicates that magnetism plays an important role. However, the specific interactions are not understood. Superconductivity in other uranium-based superconductors is often attributed to ferromagnetic spin fluctuations \cite{aoki2019review,levy2005magnetic}. This exact phenomenology cannot be applied to UTe$_2$, as UTe$_2$ is not a ferromagnet and it is not possible to identify the SC$_{\rm{FP}}$ phase with a second-order phase transition that would naturally provide enhanced spin fluctuations.

Agnostic of the underlying pairing mechanism, we can model the phase diagram of SC$_{\rm{FP}}$  phenomenologically and calculate the angle dependence of the upper critical field. Our main conclusion is that the observed angle dependence of the upper critical field is most naturally explained if the Cooper pairs carry a finite orbital momentum:

\beq
\mathbf{m}_\text{orb} = \int_{FS} \frac{\ud^2 k}{(2\pi)^2}\, \left(i \vec{d}_\kk\times \vec{d}^*_\kk\right), 
\eeq
where $\vec{d}_\kk$ is the so-called $d$-vector associated with the spin triplet order parameter. 
When the resulting orbital momentum, averaged over the Fermi surface, is non-vanishing, it will couple linearly to the magnetic field yielding a term in the free energy of the form $\Delta F = -w \mathbf{B}\cdot \mathbf{m}_\text{orb}$ which naturally depends on the field angle, as seen in the experiment.

For the orbital moment to be non-zero requires the order parameter to be multi-component and non-unitary; the latter statement reflecting the fact that the norm of the order parameter $|\Delta(\mathbf{k})|^2 \propto |\vec{d}_\mathbf{k}|^2 \mathds{1} + \mathbf{m}_\text{orb}\cdot\boldsymbol{\sigma}$ is not proportional to the identity matrix.
Moreover, for the model to account for the presence of a local maximum of the upper critical fields as a function of $\theta_a$, as well as the previously reported peak in upper critical fields in the $bc$ plane, it is important that
$\mathbf{m}_\text{orb}$ have a dominant $c$-axis component,  as described in detail in the Supplementary Materials. The likely order parameter in the SC$_\text{FP}$ phase is therefore a two-component $B_{2u} + iB_{3u}$ which has $\mathbf{m}_\text{orb}$ along the $c$-axis.

\color{black}

In summary, we present a three-dimensional mapping of SC$_{\rm{FP}}$'s angular phase boundaries and unveil its true geometry. The physical picture that emerges is a magnetically-mediated superconducting state that appears only for fields tilted off UTe$_2$'s high-field magnetic easy axis ($b$) and can be made to persist beyond 73 T. Ginzburg-Landau modeling with a non-unitary spin triplet order parameter reproduces the distinct non-monotonic angle dependence of the upper critical field.

\begin{acknowledgments}
We thank Daniel Agterberg for helpful discussions. Crystal synthesis and characterization, sample preparation, and high-field measurements were supported by the National Science Foundation under the Division of Materials Research Grant NSF-DMR 2105191. A portion of this work was performed at the National High Magnetic Field Laboratory (NHMFL), which is supported by National Science Foundation Cooperative Agreements DMR-1644779 and DMR-2128556, the State of Florida, and the Department of Energy (DOE). Electrical contact preparation was supported by the Department of Energy Award No. DE-SC-0019154. The theoretical analysis by A.H.N. was supported by the U.S. Department of Energy, Office of Basic Energy Sciences under Award no. DE-SC0025047. J.P. acknowledges support from the Gordon and Betty Moore Foundation’s EPiQS Initiative through Grant No. GBMF9071. J.S. acknowledges support from the DOE BES program ``Science of 100 T", which permitted the design and construction of much of the specialized equipment used in the high-field studies. The authors declare no competing financial interest. Identification of commercial equipment does not imply recommendation or endorsement by NIST.
\end{acknowledgments}


\bibliography{2023Halo}


\newpage

\begin{figure*}
    \centering
    \includegraphics[width=\textwidth]{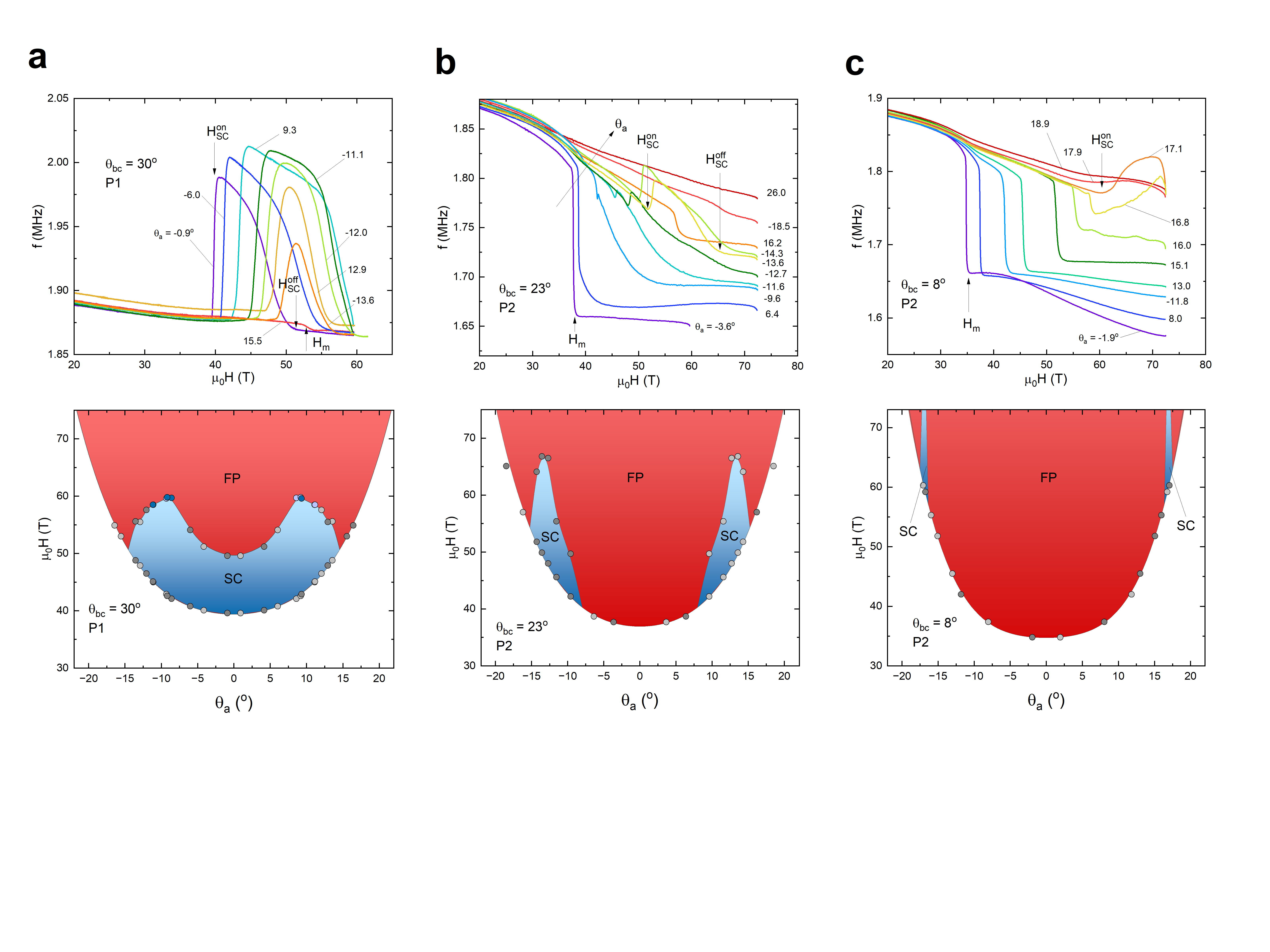}
    \caption{\textbf{Angle-dependent PDO data and corresponding phase diagrams.} PDO data (frequency vs. magnetic field) and corresponding phase diagrams collected at (a) $\theta_{bc}$ = 30$^o$, (b) 23$^o$, and (c) 8$^o$. An abrupt increase in $f$ indicates a superconducting transition while a drop indicates a metamagnetic transition into the field-polarized phase. Superconducting regions are colored blue while the electrically resistive portions of the field polarized state are colored red. The phase diagrams have been symmetrized for clarity. This is justified by the underlying crystal symmetry and the absence of any observed symmetry breaking in our measurements. Darker points indicate $\theta_a$ values at which pulses were performed while light points have been added at the corresponding -$\theta_a$ for clarity. The blue points in the phase diagram of (a) indicate that the phase transition itself was above the measured field range and was identified via extrapolation, as described in the Supplementary Materials.}
    \label{fig:2}
\end{figure*}

\newpage

\begin{figure*}
    \centering
    \includegraphics[width=\textwidth]{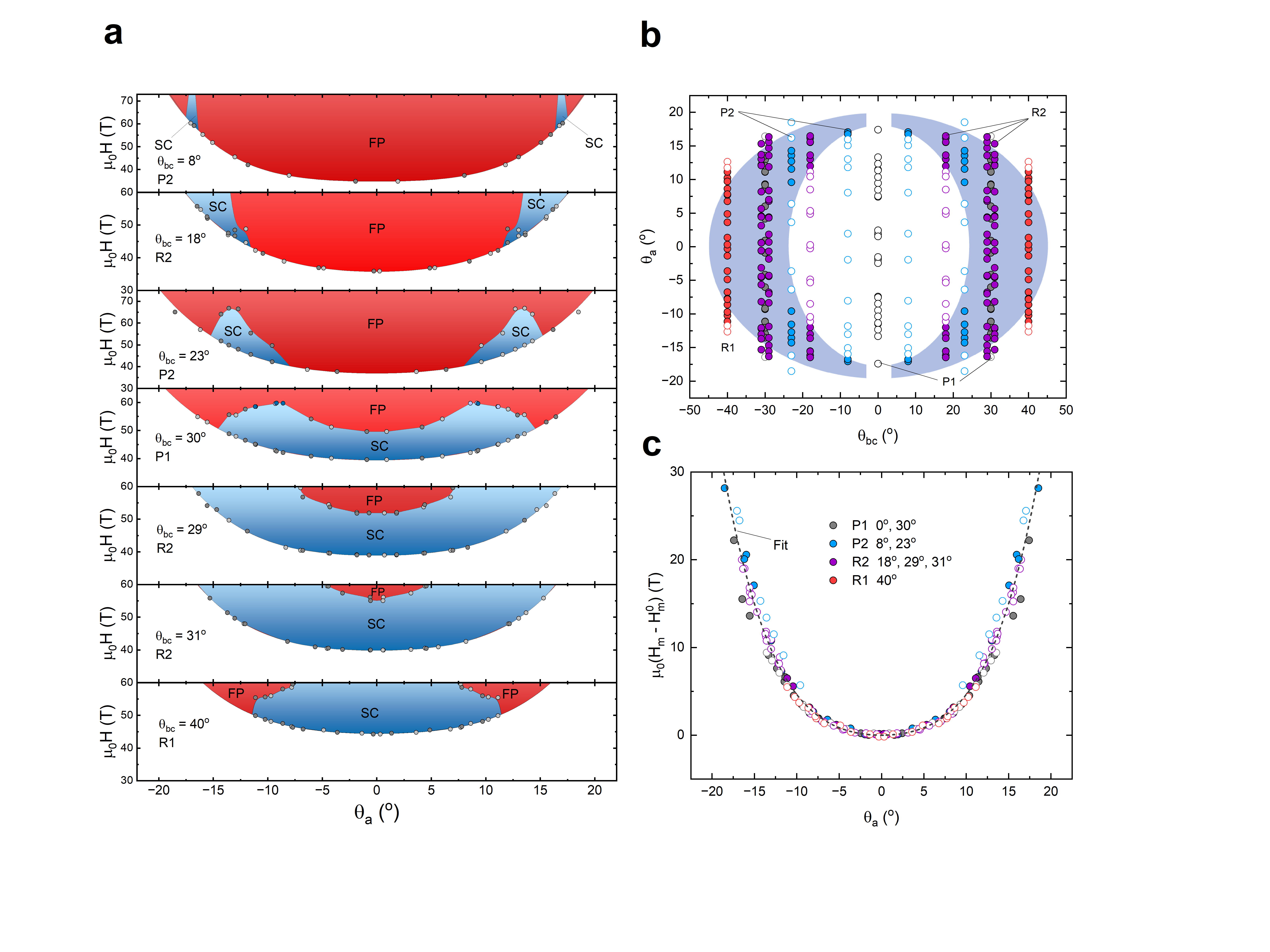}
    \caption{\textbf{Halo geometry of SC$_{\rm{FP}}$ state.} (a) High-field phase diagrams presented as a function of $\theta_a$ collected at fixed $\theta_{bc}$. Red indicates resistive regions of the FP phase while blue indicates superconducting regions. The gray points represent the measured phase boundaries from which the plots are constructed, with the dark (light) points corresponding to actual (symmetrized) data. (b) Pure angular ($\theta_{bc}$ vs. $\theta_a$) representation of the superconducting phase boundaries. Full circles are used for angles at which superconductivity is observed while open circles indicate that only a metamagnetic transition is seen. Different colors indicate different samples. Data are symmetrized to populate all four quadrants for clarity. The blue halo is for illustrative purposes, to emphasize that superconductivity is caused by tilting \textbf{H} off $b$. (c) Measured transition field minus $H_{\rm{m}}^0(\theta_{bc})$, plotted as a function of $\theta_a$. As above, full circles indicate a superconducting transition while open circles indicate that only a metamagnetic transition is seen. For a given $\theta_{bc}$, $H_{\rm{m}}^0(\theta_{bc})$ is the  metamagnetic transition field at $\theta_{a} = 0$. The legend indicates which colors are used for which sample and the $\theta_{bc}$ values at which data was collected. The dotted line is a fit of the entire data set to Eq. 1 after subtracting off the individual $H_{\rm{m}}^0 (\theta_{bc})$ values.}
    \label{fig:3}
\end{figure*}

\newpage

\clearpage

\end{document}